 \documentclass[final,5p,times,twocolumn]{elsarticle}

\usepackage{graphicx,color}
\usepackage[table]{xcolor}
\usepackage{caption}
\usepackage{subcaption}
\usepackage{amsmath}
\usepackage{epstopdf}
\usepackage{pifont}
\biboptions{numbers,sort&compress}
\newcommand{\D}{\mathrm{d}}
\newcommand{\e}{\mathrm{e}}
\newcommand{\half}{\frac{1}{2}}
\newcommand{\vecr}{\mathbf{r}}
\newcommand{\veck}{\mathbf{k}}
\newcommand{\vecrho}{\boldsymbol{\rho}}
\newcommand{\be}{\begin{equation}}
\newcommand{\ee}{\end{equation}}
\newcommand{\ba}{\begin{align}}
\newcommand{\ea}{\end{align}}
\newcommand{\bea}{\begin{eqnarray}}
\newcommand{\eea}{\end{eqnarray}}
\newcommand{\eps}{\varepsilon}
\newcommand{\dz}{\frac{\partial}{\partial z}}
\newcommand{\kbt}{k_{\mathrm{B}}T}

\newcommand{\lb}{l_\mathrm{B}}
\newcommand{\ld}{\lambda_\mathrm{D}}
\newcommand{\kd}{\kappa_\mathrm{D}}
\newcommand{\lgc}{l_\mathrm{GC}}

\newcommand{\ra}[1]{\textcolor{black}{#1} } % comments in MAGENTA

\newcommand{\kmin}{k_{\rm{min}}}
\newcommand{\qmin}{q_{\rm{min}}}

\journal{Advances in Colloid and Interface Science}

\begin{document}

\begin{frontmatter}


\title{Electrostatics of Patchy Surfaces}


\author[add1]{Ram M. Adar}
\author[add1]{David Andelman\corref{cor1}}
\ead{andelman@post.tau.ac.il}
\author[add2]{Haim Diamant}
\address[add1]{Raymond and Beverly Sackler School of Physics and Astronomy, Tel Aviv University, Ramat Aviv, Tel Aviv 69978, Israel}
\address[add2]{Raymond and Beverly Sackler School of Chemistry, Tel Aviv University, Ramat Aviv, Tel Aviv 69978, Israel}
\cortext[cor1]{Corresponding author}

\begin{abstract}
In the study of colloidal, biological and electrochemical systems, it is customary to treat surfaces, macromolecules  and electrodes as homogeneously charged. This simplified approach is proven successful in most cases, but fails to describe a wide range of heterogeneously charged surfaces commonly used in experiments. For example, recent experiments have revealed a long-range attraction between overall neutral surfaces, locally charged in a mosaic-like structure of positively and negatively charged domains (``patches"). Here we review experimental and theoretical studies addressing the stability of heterogeneously charged surfaces, their ionic strength in solution, and the interaction between two such surfaces. We focus on electrostatics, and highlight the important new physical parameters appearing in the heterogeneous case, such as the largest patch size and inter-surface charge correlations.
\end{abstract}

\begin{keyword}
Heterogeneously charged surfaces \sep ionic solutions \sep surface forces \sep hydrophobic surfaces\sep Poisson-Boltzmann theory


\end{keyword}

\end{frontmatter}

% \linenumbers
\tableofcontents
\newpage

%\baselineskip=22pt %%%%%%%%%%%%%%%%%%
%%%%%%%%%%%%%%%%%%%%%%%
\section{Introduction}
\label{introduction}
%%%%%%%%%%%%%%%%%%%%%%
Electrostatic interactions are paramount in the study of numerous colloidal, biological, and electrochemical systems. In aqueous media, some of the surface charges of macromolecules may dissociate whereas ions from the solution can bind to the macromolecules~\cite{Israelachvily,David95}. Both processes result in a net surface charge leading to a long-range Coulombic interaction, mediated by ions in the solution. The interplay between electrostatics and the ion entropy of mixing is described by the Poisson-Boltzmann (PB) theory. Combining the universal van der Waals (vdW) interaction with PB theory yields the well-known Derjaguin-Landau-Verwey-Overbeek (DLVO) theory~\cite{DL,VO}.

Traditionally, DLVO theory is used to study the interaction between homogeneously charged surfaces~\cite{Ohshima,Safinyabook}. Homogeneity is an idealization, as charges are distributed discretely on the molecular level, and surfaces can also be heterogeneously charged over mesoscopic length scales (nanometers to micrometers), either spontaneously or by design~\cite{Israelachvily,Ohshima}. The latter has stimulated many experimental and theoretical works in the past few decades, addressing surface-charge heterogeneity on microscopic and mesoscopic levels, under a wide range of physical conditions.  Different aspects of inhomogeneity have been investigated, including the stability of surface-charge heterogeneities, counterion distribution at the surface proximity, and interactions between two such surfaces across an ionic solution.

The study of heterogeneously charged (``patchy") surfaces has gained a growing interest during the last decade due to novel experiments, which measured a long-range attraction between hydrophobic surfaces across an aqueous solution~\cite{Perkin05,Meyer05,Hammer10,Perkin06,Meyer06,SilbertPRL}. Although these neutral surfaces were initially homogeneous during preparation, it was observed that they transform into mosaic-like structures of positively and negatively charged patches. These patchy surfaces remain stable during experimental times, and it has been established ~\cite{Meyer05,Hammer10,Perkin06,Meyer06,SilbertPRL} that the measured long-range ``hydrophobic" attraction was in fact of electrostatic origin.

In the present work, we review experimental and theoretical works concerned with the electrostatic properties of patchy surfaces in an ionic solution. We highlight the important features of surface-charge heterogeneity, while describing some of the theoretical frameworks related to surface-force experiments.

The outline of this paper is as follows. In the next section, we focus on the experimental aspects of patchy surfaces, describing their preparation and the forces measured between them. In Section~\ref{sec3}, we discuss under what conditions finite-size charged patches are formed and what is their expected optimal size in solution. In Section~\ref{sec4}, we turn to ionic profiles in the proximity of such patchy surfaces for different ionic environments. Next, we discuss  in Section~\ref{sec5} the osmotic pressure between two heterogeneously charged surfaces and relate it to the measured long-range attraction between patchy surfaces. Finally, in Section~\ref{sec6}, we compare this long-range attraction with the ever-present van der Waals attraction.

%%%%%%%%%%%%%%%%%%%%%%%
\section{Patchy surfaces in experiments}
\label{sec2}
%%%%%%%%%%%%%%%%%%%%%%

\subsection{Preparation of patchy surfaces}
\label{preparation}
Patchy surfaces can be prepared by different methods. We focus here on the methods used separately by the groups of Israelachvili~\cite{Meyer05,Meyer06,Hammer10} and Klein~\cite{Perkin05,Perkin06,SilbertPRL}, whose experiments inspired a number of works, as discussed below. In these experiments, the patchy surfaces consist of a positively charged bilayer of surfactants adsorbed on a negatively charged mica surface. During preparation, the anionic mica surface is first coated  with a cationic surfactant monolayer by self-assembly of cetyltrimethylammonium bromide or fluoride (CTAB or CTAF, respectively) from the aqueous solution~\cite{Perkin05,Perkin06,Meyer06,SilbertPRL}. Another technique is the  Langmuir-Blodgett (LB) method~\cite{Meyer05,Meyer06}, in which surfactants such as dimethyldioctadecylammonium bromide (DODAB) are used. By adsorbing surfactants with hydrophobic tails, the mica surface itself becomes hydrophobic, and such surfaces have been used over the years in many experiments studying hydrophobic surfaces~\cite{Christenson01,Christenson88,Wood95,Lin05,Chen92,Pashley81,Eriksson97}.

\begin{figure}[ht]
\centering
\includegraphics[width=1\columnwidth]{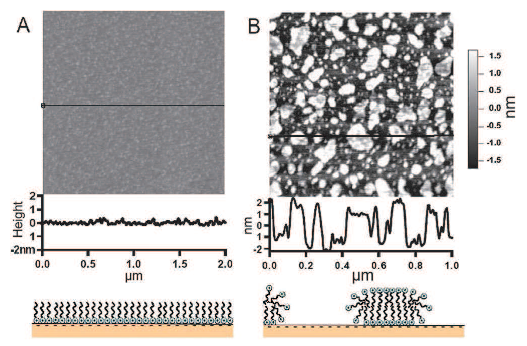}
\caption{(Color online) AFM images of a surfactant-coated mica surface. (a) In a monolayer structure, immediately after deposition,
and (b) in a bilayer structure after immersion time of 22 hours in pure water. The cartoons illustrate the breakup of the initially uniform monolayer into positively charged bilayer domains on an otherwise negatively charged mica substrate. Reproduced with permission from Ref.~\cite{SilbertPRL}. }
\label{fig1}
\end{figure}

 However, the monolayer structure is less favorable in solution than that of a bilayer, for which the hydrophobic surfactant tails are confined and only their cationic heads are in contact with the polar water molecules. The inner surfactant monolayer then neutralizes the negative mica, while the outer monolayer makes the bilayer domains positive, as is illustrated in Fig.~\ref{fig1}.
We note that different methods have been used to prepare patchy surfaces also for silica~\cite{Zhang05} and latex~\cite{Popa10}.

 Three experimental findings indicate that the surfactant monolayer structure indeed breaks into patches of bilayers over timescales of a few hours. First, contact angle measurements show that the coated mica surface becomes less hydrophobic in solution within a few hours~\cite{Perkin05,Perkin06}, implying surfactant disassociation. Second, the inter-surface separation at which the two mica sheets jump into contact, increases with time from twice the thickness of a monolayer to four times that thickness~\cite{Perkin06}. This indicates a transition from a monolayer-monolayer contact to a bilayer-bilayer one. The third and most direct observation comes from atomic force microscopy (AFM) images. The microscope measures force curves that can be fitted to the DLVO predictions \cite{Butt92,Rotsch97,Heinz99,Yin08,Drelich10,Drelich11}. By moving the probe across the substrate, it is then possible to map the local electrostatic forces and charge density, capturing the bilayer structure itself. AFM images of surfactant-coated mica are shown in Fig.~\ref{fig1}.

 %%%%%%%%%figure%%%%%%%%%%%%%%%
%%%%%%%%%%%%%%%%%%%%%%%%%%%%%%%

\subsection{Measured long-range attraction between patchy surfaces}
\label{patchy_forces}

\begin{figure}[ht]
\centering
\includegraphics[width=0.85\columnwidth]{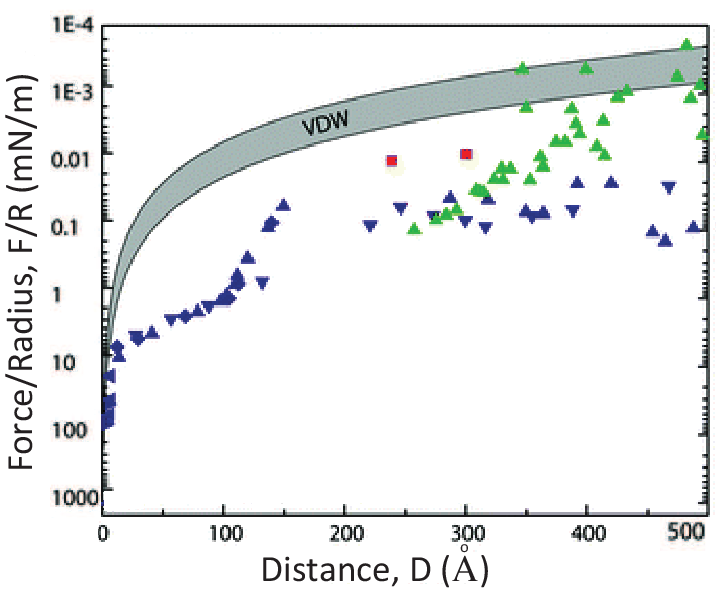}
\caption{(Color online) Measured attractive forces (in absolute value) between coated mica surfaces as a function of the inter-surface distance. The force is normalized by the surface radius of curvature, $R$, and plotted on a logarithmic scale. The symbols correspond to different coating materials and methods. Red squares correspond to LB-deposited DMDOA, green triangles correspond to chemical-vapor-deposited OTE, and blue triangles and inverted triangles correspond to LB-deposited OTE and DODA surfaces, respectively. The red squares and green triangles were measured after deaeration. The shaded VDW force band corresponds to Hamaker constant values between $3$ and $10\times10^{-21}\,\rm{J}$. The different experiments from which the data points are taken are detailed in Ref.~\cite{Hammer10}. Reproduced with permission from Ref.~\cite{Hammer10}.}
\label{fig2}
\end{figure}

Force measurements between patchy surfaces were first performed without prior knowledge of any patchy structure.  As is explained in Section~\ref{preparation}, patchy surfaces can form spontaneously from surfactant-coated mica surfaces. Such surfaces are initially hydrophobic and were used~\cite{Christenson01,Christenson88,Wood95,Lin05,Chen92,Pashley81,Eriksson97} to investigate the {\it hydrophobic interaction}, referring to the attraction between non-polar surfaces across water. Measurements were conducted under a wide range of experimental conditions, revealing long-range attractive forces at distances ranging up to hundreds of nanometers~\cite{Meyer06}. These distances are much larger than the typical range of vdW attraction, as is demonstrated in Fig.~\ref{fig2}.

\ra{Several mechanisms have been suggested as an explanation for the attraction. While the genuine hydrophobic interaction is considered to occur only for separations smaller than 20 nm, the long-range attraction is mainly attributed to Coulombic interactions~\cite{Hammer10}. For example, Podgornik and Parsegian~\cite{Rudi9195} related the attraction to an instability in the amount of adsorbed surfactants, resulting from the changes in the inter-surface separation.}

%%%%%%%%%% patches forces %%%%%%%%%%%
%%%%%%%%%%%%%%%%%%%%%%%%%%%%

The understanding of the long-range force and its relation to the hydrophobic nature of the surfaces has been advanced by Meyer et al.~\cite{Meyer05,Meyer06} and Perkin et al.~\cite{Perkin05,Perkin06}. Using the surface force apparatus (SFA), both groups demonstrated that the interaction is screened by added salt, implying its electrostatic origin. Moreover, it was shown that the surfactant monolayer coating the mica dissociates and forms patchy bilayers, as is explained in Section~\ref{preparation}. This indicates that the long-range attraction is indeed electrostatic, and can be attributed to  {\it annealing} of surface patches. Assuming relaxation times shorter than the measurement times, charged patches on one surface are free to rearrange and position themselves against oppositely charged patches on the other surface and vice versa. Consequently, the surfaces are composed of correlated oppositely charged patches, and the attraction can be understood by means of simple electrostatics. This scenario is particularly important when the surfaces are in close contact.

However, one can also consider another scenario of {\it quenched} systems, where the relaxation time is longer than experimental times, and the patch arrangement is effectively frozen in time. The quenched scenario was tested in the work of Silbert et al. \cite{SilbertPRL}, who applied an in-plane velocity on one of the surfaces while measuring the inter-surface force, frustrating any patch rearrangement on experimental time scales. Remarkably, the attraction prevails and retains its magnitude. This result is counter-intuitive because overall neutral surfaces without any correlations are not expected to exhibit electrostatic interaction on average, leaving a predominant entropic repulsion of the mobile ions. The differences between the annealed and quenched scenarios are further investigated in Section~\ref{sec5}.

%%%%%%%%%%%%%%%%%%%%%%%%%%%%%%%%%%%%%%%%%%%%%%%%%%%%%%%%%%%%%%%%%
\section{Modeling of patch formation and optimal patch size}
\label{sec3}
%%%%%%%%%%%%%%%%%%%%%%%%%%%%%%%%%%%%%%%%%%%%%%%%%%%%%%%%%%%%%%%%%

  The stability of patchy mica sheets described above depends on several parameters, such as the immersion time of mica in the surfactant solution during the initial coating process and surfactant composition~\cite{Perkin05}, as well as pH~\cite{Chen92} and the ionic strength of the solution~\cite{Eriksson97}. Rather than describe the experimental setup, we change gears and present a simple model of mobile (annealed) positive and negative charges on a neutral surface.

\begin{figure}[ht]
\centering
\includegraphics[width=0.9\columnwidth]{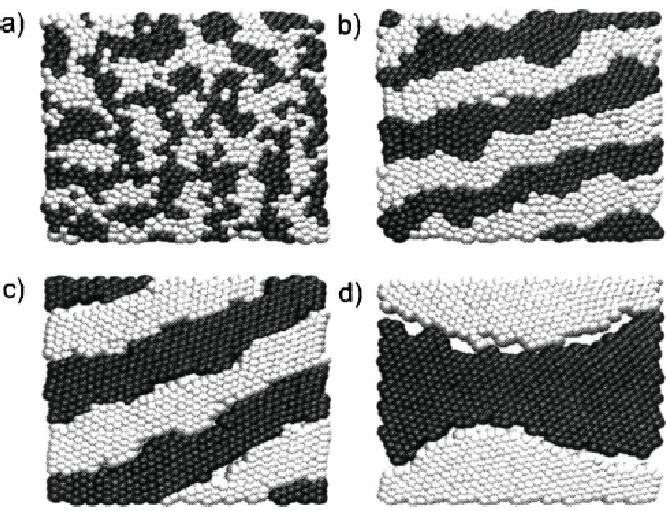}
\caption{Molecular dynamics  (MD) snapshots of two mobile and oppositely charged species (marked black and white) confined to a surface in contact with an aqueous solution. The repulsive interaction between the two species increases from (a) to (c), resulting in larger domains. In (d), salt has been added to the solution,  screening the electrostatics and leading to phase separation. Reproduced with permission from Ref.~\cite{Loverde06}.}
\label{fig3}
\end{figure}

  Generally, a binary mixture of oppositely charged species may not always exhibit a stable structure of finite-size patches. While electrostatics promotes charge mixing, short-range interactions can induce phase separation into macroscopically large domains. The interplay between the two mechanisms, therefore, determines whether patches are stable and what would be their optimal size. An elucidating description of this interplay is given in the works of  de la Cruz and co-workers~\cite{Solis05,Velichko05,Loverde06,Loverde07,Solis11}, and Pincus and Safran and co-workers~\cite{Naydenov07,Brewster08,Jho11}. In their works, it was shown that charged surface domains are stable as long as the Coulombic interaction is sufficiently strong. Once the interaction is screened, the domain size increases and, finally, a first-order transition takes place. This is illustrated in Fig.~\ref{fig3} and is described in more detail below.
%%%%%%%%%%%%%%%%%%%%%%%%%%%%

Consider two oppositely charged and mobile species, distributed periodically on \ra{an infinitely rigid} surface composed of unit cells of area $A=l^{2}$. The free energy per unit cell area, $f=F/(NA)=f_{\rm{el}}+f_{\rm{SR}}$, where $N$ is the number of unit cells, consists of long-range electrostatics, $f_{\rm{el}}$, and short-range interactions, $f_{\rm{SR}}$. The former scales as $\sigma^{2}l/\eps$, where $|\sigma|$ is the absolute value of the average surface-charge density, and $\eps$ is the dielectric constant of the solution. The term $f_{\rm{SR}}$ scales as $\tau/l$, where $\tau$ is the line tension (energy per unit length).  Combining the two, we find that~\cite{Solis05}
%%%%%%%%%eq%%%%%%%%%%%
\be
\label{eq5}
f=s_{1}\frac{\tau}{l}+s_{2}\frac{\sigma^{2}l}{\eps},
\ee
%%%%%%%%%%%%%%%%%%%%%%%
where the dimensionless coefficient $s_{1}$ is the ratio between the unit-cell contour length and the cell size $l$, and $s_{2}$ corresponds to the average electrostatic free energy per unit cell. Ignoring numerical prefactors and equating the two terms, one finds a characteristic length scale $l_{0}=\sqrt{\tau\eps/\sigma^{2}}$, which defines a characteristic energy (per unit area) $f_{0}=\sqrt{\tau\sigma^{2}/\eps}$. Rescaling Eq.~(\ref{eq5}) yields
%%%%%%%%%%%%%%%eq%%%%%%%%%%%%
\be
\label{eq1}
\frac{f}{f_{0}}=s_{1}\frac{l_{0}}{l}+s_{2}\frac{l}{l_{0}}.
\ee
%%%%%%%%%%%%%%%%%%%%%%%%%%%%%%%

For Coulombic interactions, as $s_{2}$ is of order unity, minimization of the free energy leads to a finite domain size, $l^{\ast}\sim l_{0}$~\cite{Solis05,Naydenov07}.  When salt is added, the electrostatics becomes screened, and $s_{2}l/l_{0}$ becomes of order unity. The free energy is then minimal for $l^{\ast}/l_{0}\to\infty$, corresponding to a phase separation. In the case of high salt concentrations, $s_{2}$ can be calculated using the linear Debye-H{\"u}ckel (DH) framework, which is applicable for small electrostatic potentials, $\psi<25\,\rm{mV}$~\cite{David95}. Within the DH framework, charges separated a distance $r$ apart interact via a Yukawa-like potential $\sim\exp\left(-\kd r\right)/r$, where $\kd^{-1}=\ld$ is the inverse Debye screening length. For ions of valencies $z_{i}$ and bulk concentrations $n^{0}_{i}$, $\kd$ is given by
%%%%%%%%%%%eq%%%%%%%%%%
\be
\label{eq5b}
\kd=\sqrt{4\pi \lb\sum_{i}z_{i}^{2}n^{0}_{i}},
\ee
%%%%%%%%%%%%%%%%%%%%%%%
and for monovalent salt ($z_{i}=\pm1$) of bulk concentration $n_{0}$, it simplifies to $\kd=\sqrt{8\pi\lb n_{0}}$. The Bjerrum length, $\lb$, is defined as $\lb=e^{2}/\left(\eps\kbt\right)$ \ra{(Gaussian units)}. For water, $\eps\approx 78$, and at room temperature, $\lb\approx0.7\,{\rm nm}$. Another important legnthscale is the Gouy-Chapman length, $\lgc=e/\left(2\pi\lb|\sigma|\right)$, at which the Coulombic interaction of a homogeneously charged surface of charge density $\sigma$ with an elementary charge is equal to the thermal energy.

As an example for calculating $s_{2}$ with DH theory, we consider a stripe structure of 1D domains where the surface-charge density, $\sigma$, is described by a single sinusoidal mode, $\sigma(x)=\sigma_{k}\sin(kx)$, where $k=2\pi/l$ is the modulation wavenumber. One finds~\cite{Naydenov07} that $s_{2}\propto k/\sqrt{k^{2}+\kd^{2}}$, and  by minimizing the free energy, the optimal wavenumber, $k^{\ast}$, is obtained. Further analysis indicates that a first-order phase transition between stable finite domains and a macroscopic phase separation occurs for $\kd\l_{0}\approx1$~\cite{Naydenov07}.

\begin{figure}[ht]
\centering
\includegraphics[width=0.7\columnwidth]{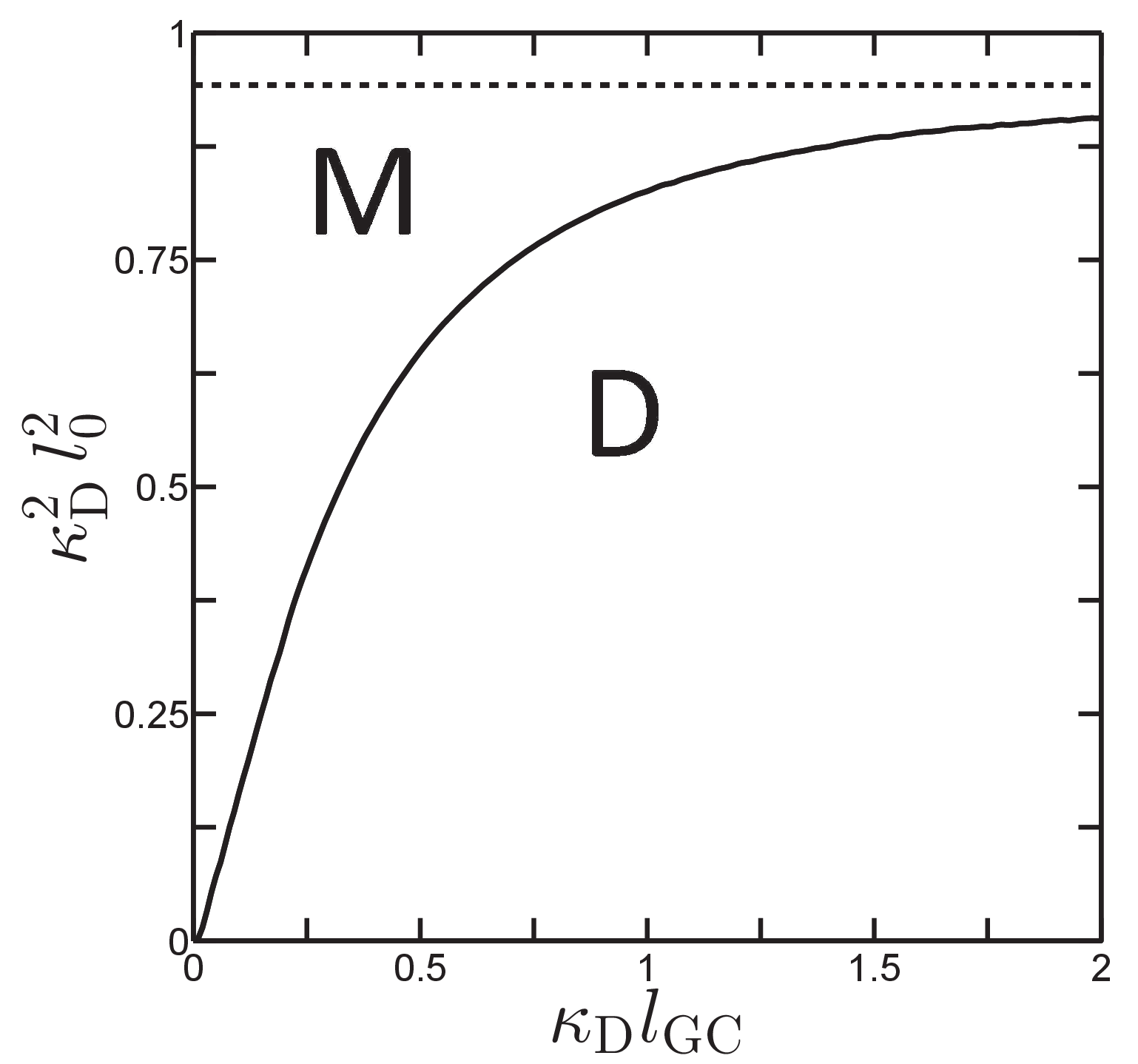}
\caption{Phase diagram of a surface composed of two oppositely charged species, either forming finite domains (D) or undergoing macroscopic phase separation (M). The result from a variational approach is plotted as a solid curve and the DH result as a dashed horizontal line. Adapted from Ref.~\cite{Naydenov07}.}
\label{fig4}
\end{figure}

 At intermediate salt concentrations, the calculation of $s_{2}$ cannot be done analytically, and often simulations are used~\cite{Velichko05,Loverde06,Loverde07}. The optimal wavenumber, $k^{\ast}$, is then defined as the value for which the 2D structure factor is at its peak. Naydenov et al.~\cite{Naydenov07} presented a variational approach to this regime, minimizing the PB free energy with respect to a variational ansatz for an electrostatic potential of the form $\psi(x,z)=\sin(kx)h(z)$. The phase diagram they obtained within this framework and within the DH approximation is presented in Fig.~\ref{fig4}.

The phase transition between finite-size domains and macroscopic phase separation has been studied in more general setups, e.g., for cylinders~\cite{Solis05,Velichko05,Solis11}. Furthermore, for two surfaces interacting across a solution, it was shown that a phase transition occurs even without salt~\cite{Brewster08}. Examining a two-surface system, the free energy is lowest when the surfaces are oppositely charged, having opposite contributions to the electric field. As the inter-surface separation, $d$, becomes smaller, these competing contributions result in a diminished electric field. In this effective screening mechanism, $d^{-1}$ plays the role of $\kd$, and the electrostatics becomes sufficiently weak for phase separation to take place.
%%%%%%%%%%%%%%%%%%%%%%%%%%%%
%%%%%%%%%%%%%%%%%%%%%%%%%%%%%%%%%%%%%%%%%%%%%%%%%%%%%%%%%%%%%%%%%

\section{Ionic profiles near a single patchy surface}
\label{sec4}
%%%%%%%%%%%%%%%%%%%%%%%%%%%%%%%%%%%%%%%%%%%%%%%%%%%%%%%%%%%%%%%%%
A heterogeneous fixed surface-charge density results in a heterogeneous charge density of mobile ions in solution. Although the charge density can be determined by the electrostatic potential, $\psi$, via Poisson's equation, finding the electrostatic potential proves to be a challenging task. It is common to assume that the ions obey a Boltzmann distribution, determined by the average electrostatic potential induced by the fixed charges and all other ions. Together with Poisson's equation, this assumption leads to the mean field (MF) Poisson-Boltzmann (PB) equation:
%%%%%%%%%eq%%%%%%%%%
\be
\label{eq5a}
\eps\nabla^{2}\psi(\vecr)=-4\pi\left(\rho_{f}(\vecr)+e\sum_{i}z_{i}n^{0}_{i}\e^{-ez_{i}\psi(\vecr)/\kbt}\right),
\ee
%%%%%%%%%%%%%%%%%%%
where $\rho_{f}$ is the charge density of fixed surfaces and macromolecules and $n^{0}_{i}$ is the reference ionic number density of the $i$th species, which coincides with the bulk value. The DH theory is then obtained by Taylor-expanding the exponent in Eq.~(\ref{eq5a}) up to first order in the potential.

The PB framework has its own limits. Due to a combination of large fixed charge densities and high ionic valencies, strong electrostatic correlations can become important, rendering the MF formulation inadequate. This regime is referred to as the {\it strong coupling} (SC) regime~\cite{Moreira00,Moreira01,Netz01}. In what follows, we discuss three different scenarios, including the SC one: the added-salt and counterions only cases within PB, and the SC case. The major differences between the three are summarized in Table ~\ref{table1}.
%%%%%table for different cases%%%%
\begin{table}[ht]
\centering
\begin{tabular}{|l|c|c|c|c|}

   \hline
   \cellcolor[gray]{0.9}& PB & Correlations & Co-ions & DH limit \\ \hline
  Counterion-only & \ding{51} & \ding{55}& \ding{55} & \ding{55} \\ \hline
  Added salt & \ding{51} & \ding{55} & \ding{51} & \ding{51} \\ \hline
  SC & \ding{55} & \ding{51} & \ding{55} & \ding{55} \\ \hline
  \end{tabular}
  \caption{Three electrostatic models as discussed in the paper. The counterion-only and added salt cases are described within PB theory, while the SC regime corresponds to systems with highly correlated counterions whose properties cannot be captured using MF theory. Co-ions are present only in the added salt systems, for which the DH limit can be used for small electrostatic potentials.}
  \label{table1}
  \end{table}
%%%%%%%%%

\begin{figure}[ht]
\centering
\includegraphics[width=0.5\textwidth]{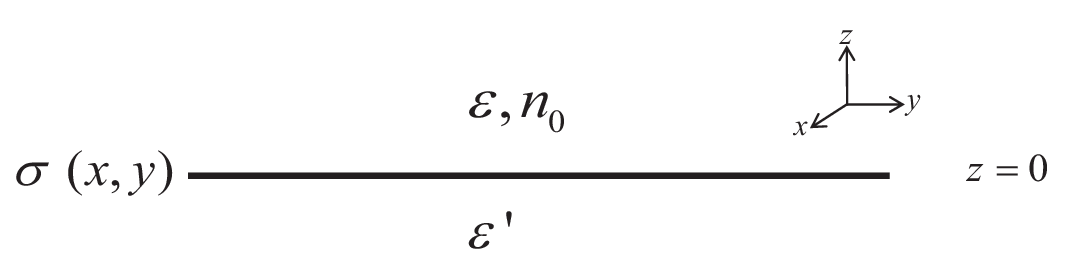}
\caption{Schematic drawing of a flat surface at $z=0$, with charge density  $\sigma(x,y)$. The dielectric
constant in the upper and lower regions is $\eps$ and $\eps'$, respectively. The aqueous solution in the upper region is in contact with a monovalent salt reservoir of concentration $n_{0}$. }
\label{fig5}
\end{figure}

In order to investigate the ionic profiles near patchy surfaces, we consider the following electrostatic setup, as is depicted in Fig.~\ref{fig5}: a planar surface at $z=0$ is charged heterogeneously with a surface-charge density $\sigma\,(x,y)$. The surface area, $A$, is taken to be macroscopic, $A\to\infty$, and separates an ionic solution at $z>0$ from a homogenous dielectric at $z<0$. The ionic solution has a dielectric constant $\eps$ and a monovalent salt reservoir of concentration $n_{0}$, while the medium at $z<0$ has a dielectric constant $\eps'$.
%%%%%%%%%%%%%%%%%%%%%%%%%%%%%%%%%%%%%%%%%%%%%%%%%%%%%%%%%%%
%%%%%%%%%%%%%%%%%%%%%%%%%%%%%%%%%%%%%%%%%%%%%%%%%%%%%%%%%%%
\subsection{Added salt: DH theory}
\label{DH_concentrations}
 The nonlinear PB equation [Eq.~(\ref{eq5a})] cannot be solved analytically for heterogeneously charged bounding surfaces, and some approximations are required. For added salt, the most straightforward approximation is the linear DH theory, described in Section~\ref{sec3}. Below, we present the DH results for the ionic concentration near one heterogeneously charged surface. This case will serve as a basis for the remainder of this section.

 In the electrostatic setup of Fig.~\ref{fig5}, the potential solves Laplace's equation for $z<0$, while for $z>0$, it solves the DH linear differential equation, {\it i.e.,}
 %%%%%%%%%%%eq%%%%%%%%%
 \be
 \label{eq1b}
 \nabla^{2}\psi=\begin{cases}
0 & z<0\\
\kd^{2}\psi & z>0.
\end{cases}
 \ee
 %%%%%%%%%%%%%
 Here $\kd$ is given by  $\kd=\sqrt{8\pi\lb n_{0}}$.  The boundary condition for the electrostatic potential at $z=0$ is given by the surface-charge density,
 %%%%%%%%%%%%eq%%%%%%%%%%%%%
 \be
 \label{eq5d}
 \left.\eps\dz\psi\right|_{0^{+}}-\left.\eps'\dz\psi\right|_{0^{-}}=-4\pi\sigma\,(x,y).
 \ee
 %%%%%%%%%%%%%%%%%%%%%%%%%%%%%%%%
  The excess number density of positive and negative ions, $\Delta n^{\pm}(\vecr)=n^{\pm}(\vecr)-n_{0}$, can be conveniently expressed via a Fourier transform in the in-plane coordinates. Defining $\vecrho=(x,y)$ and the 2D Fourier transforms,
 %%%%%%%%%%%eq%%%%%%%%%%%%
 \begin{align}
 \label{eq5e}
 \Delta n^{\pm}_{\veck}(z)&=\int \Delta n^{\pm}(\vecrho,z)\e^{-i\veck\cdot\vecrho}\,\D^{2}\rho,\nonumber\\ \sigma_{\veck}&=\int\sigma(\vecrho)\e^{-i\veck\cdot\vecrho}\,\D^{2}\rho,
 \end{align}
%%%%%%%%%%eq%%%%%%%%%%%%%%%%%%
one finds for $z>0$
%%%%%%%%%%%%%%%%%%%%%%%%%%%%%
\begin{align}
\label{eq7}
\Delta n^{\pm}_{\veck}(z)&=\mp\frac{1}{2}\eps\kd^{2}\frac{\sigma_{\veck}}{e}\frac{\e^{-qz}}{q\eps+k\eps'}\,,
\end{align}
%%%%%%%%%%%%%%%%%%%%%%%%%%%%%%
where $q=\sqrt{\kd^{2}+k^{2}}$. In the homogeneous case,  the ionic densities decay as $\exp\left(-\kd z\right)$. Here, each $k$-mode decays with a modified inverse screening length, $q$, combining the contributions of salt (via $\kd$) and the $k$-mode of alternating positive and negative surface charges. Consequently, higher modes (smaller wavelengths) decay faster. This implies that at very large separations, the ionic profiles are determined solely by the lowest $k$-mode, $\kmin$. This mode decays as $\exp\left(-\qmin z\right)$, where $\qmin^{2}=\kd^2+\kmin^2$, and corresponds to the largest patch size on the surface.

 Hereafter, unless mentioned otherwise, we assume that the electric field is confined to the aqueous solution ($z\ge0$), and is zero for $z<0$.  Given the high dielectric constant of water, $\eps\approx80$, many systems exhibit a large dielectric mismatch, $\eps\gg\eps'$, justifying this assumption. Furthermore, interfaces of physical systems have a finite thickness, along which the electric field diminishes. For example, the ``oily" part of a membrane has a dielectric constant $\eps'\approx2$, and the above assumption is considered to hold for $h/\ld\gg\eps'/\eps\approx1/40$, where $h$ is the membrane thickness~\cite{Kiometzis89,Winterhalter92}.

\subsection{Counterion-only case}
\label{WC_concentration}

Within the DH framework, heterogeneously charged surfaces induce spatially-dependent ionic concentrations, even when the surfaces are overall neutral. In the counterion-only scenario, on the other hand, there are no counterions in the absence of net charge. This result can also be obtained by substituting $\kd=0$ in Eq.~(\ref{eq7}). Screening still occurs due to the surface-charge modulations, as is evident by solving the DH equation that reduces to Laplace's equation for $\kd=0$.

In the case where there is a net surface-charge and no added electrolyte, other approximations of the full PB equation [Eq.~(\ref{eq5a})] are in order. One of them is to expand the dimensionless electrostatic potential, $\phi=e\psi/\kbt$, in the PB equation  perturbatively with respect to a small parameter $\epsilon$, which relates to the surface-charge inhomogeneity, {\it i.e.}, $\phi=\phi_{0}+\epsilon\phi_{1}+\epsilon^{2}\phi_{2}+\mathcal{O}\left(\epsilon^{3}\right)$. The term $\phi_{0}$ is the potential for a homogeneously charged surface bearing the same net charge. Higher-order terms depend on lower ones and can be iteratively solved. This scheme is sometimes referred to as partially/modified linearized PB~\cite{Milkavcic95,White02}. Note that $\phi_{0}(z)$ solves the regular PB equation, while $\phi_{1}(\vecr)$ \ra{and $\phi_{2}(\vecr)$}, for example, solve the equations
 %%%%%%%%%%eq%%%%%%%%%%%%%%%%%%%%%%
 \begin{align}
 \label{eq8}
 \nabla^{2}\phi_{1}&=4\pi\lb n_{0}\e^{-\phi_{0}}\phi_{1},\nonumber\\
 \nabla^{2}\phi_{2}&=4\pi\lb n_{0}\e^{-\phi_{0}}\left(\phi_{2}-\half \phi_{1}^2\right).
 \end{align}
 %%%%%%%%%%%%%%%%%%%%%%%%%%%%%%%%%%
In the counterion-only scenario, there are no ions in the bulk, and the reference number-density, $n_{0}$, is determined by the electro-neutrality condition (Gauss' law). This method can also be  used to solve the potential beyond the DH framework for the case of added salt.

\ra{As $\phi_{2}$ in Eq.~(\ref{eq8}) depends on $\phi_{1}^2$, the contributions coming from the different $k$-modes of the surface-charge density [Eq.~(\ref{eq5e})] are coupled. The same holds for higher orders, $n>2$.} In particular, $k>0$ modes affect the $k=0$ mode of the ionic density, thus modifying the area-averaged ionic density, as was demonstrated by Lukatsky et al.~\cite{Lukatsky02,Lukatsky02b}. By solving Eq.~(\ref{eq8}) and from Monte-Carlo (MC) simulations, it was shown that for overall charged surfaces, surface-charge modulations enhance the counterion contact density (at the surface proximity). Namely, the counterion density satisfies
 %%%%%%%%eq%%%%%%%%%
 \be
 \label{eq9}
\frac{\langle n\left(\vecrho,0\right)\rangle_{\vecrho}}{n_{h}(0)}=1+\int\frac{\D^{2}k}{(2\pi)^{2}}\,w(k\lgc)\left|\frac{\sigma_{\veck}}{\sigma_{0}}\right|^{2},
 \ee
 %%%%%%%%%%%%%%%%%%%
where $\langle n(\vecrho,z)\rangle_{\vecrho}$ is the area-averaged counterion density and $n_{h}(z)=1/\left[2\pi\lb\left(z+\lgc\right)^{2}\right]$ is the counterion density profile for a homogeneously charged surface bearing the same net charge. The weight function, $w(x)$, is the calculated contribution of the $k$-th mode to the counterion density (not shown here). Note that $\lgc$ is defined with respect to the average surface-charge density, $\lgc=e/\left(2\pi\lb|\sigma_{0}|\right)$.

 The function $w(k\lgc)$ in Eq.~(\ref{eq9}) is positive and monotonically decreasing, satisfying $w(0)=1$. These properties convey the important features of the calculation. As the function is positive, the increase in counterion density at the surface is a global effect, independent of the exact form of the surface-charge modulation. Because it is a monotonically decreasing function, the effect of smaller modes is more evident, similar to the DH case. The value $w(0)=1$ implies that in the limit of large wavelengths of the modulation and/or large net charge, the magnitude of the effect is determined solely by the value of the integral $(2\pi)^{-2}\int\D^{2}k\,\left|\sigma_{\veck}/\sigma_{0}\right|^{2}=\int\D^{2}\rho\left[\sigma(\vecrho)/\sigma_{0}\right]^{2}$ (Parseval's identity).

 The localization of counterions near an overall charged surface was demonstrated in other setups. For surfaces with random discrete charges, correlations beyond MF were shown to increase the counterion concentration near the surface via a loop expansion of the system's free energy~\cite{Fleck05}. In such a setup, a similar effect can be captured also in MF, within a possible charge regulation process. Counterions can bind to the surface in order to decrease its effective charge density in absolute value, thus becoming localized in its vicinity~\cite{Henle04}.

 Several studies have investigated the counterion distribution in the presence of surface-charge inhomogeneities in both cylindrical~\cite{Landy10} and spherical \cite{Allahyarov01,Messina01,Messina02,JoePPM} geometries. A detailed discussion of these non-planar results lies beyond the scope of the present paper.
%find more works
\subsection{Strong coupling (SC) regime}
\label{SC_concentration}

Within the SC framework, counterions strongly repel each other and/or are strongly attracted to the surface, resulting in relatively separated ions positioned close to the surface. Consequently, the system properties can be determined to some extent by those of a system with a single counterion~\cite{Moreira00,Moreira01,Netz01}.  It is possible to distinguish between this framework and the PB one via the electrostatic coupling parameter, $\Xi=z^{3}\lb/\lgc$, where $z$ is the ion valency.  The  PB regime is relevant for $\Xi\ll1$, while the SC regime for $\Xi\gg1$.

The above picture holds only for mobile charges of the same sign, and the following discussion is restricted to the infinite dilution limit, where only counterions are present. In addition, the surface charges are all of only one sign, exhibiting surface-charge inhomogeneity of the form $\sigma(\vecrho)=\sigma_{0}+\sigma_{1}(\vecrho)$, where $\sigma_{0}$ is the average surface-charge density and the modulation around it satisfies $\langle \sigma_{1} (\vecrho)\rangle_{\vecrho}=0$, such that both $\sigma_{0}$ and $\sigma_{0}+\sigma_{1}$ are positive.  Solutions containing both positive and negative ions require another framework, e.g., the dressed counterion theory \cite{Kanduc10}, and lies outside the scope of this paper.

The SC counterion concentration is given by the Boltzmann distribution of a single counterion in an external potential, $n(\vecrho,z)=b\exp\left(-u(\vecrho,z)\right)$ \cite{Moreira00,Moreira01,Netz01}, where $u$ is the dimensionless electrostatic interaction energy between the surface and a single counterion. The parameter $b$ is determined by the electro-neutrality condition,
 %%%%%%%%%%%%%eq%%%%%%%%%%
 \be
 \label{eq10a}
 \int\D ^{2}\rho\int_{0}^{\infty}\D z\, e n(\vecrho,z)=\sigma_{0}A.
\ee
%%%%%%%%%%%%%%%%%%%%%%%%%%%
As the interaction energy depends linearly on the surface-charge density, it is possible to decompose it into two terms, $u=u_{0}+u_{1}$, stemming, respectively, from $\sigma_{0}$ and $\sigma_{1}$. The counterion concentration, $n\left(\vecrho,z\right)$, can then be written as
%%%%%%%%%%%%%%%%eq%%%%%%%%%%%%%%%%%%
\be
\label{eq10}
n(\vecrho,z)=\Lambda n_{h}(z)\e^{-u_{1}(\vecrho,z)},
\ee
%%%%%%%%%%%%%%%%%%%%%%%%%%%%%%%%%%%%
where $\Lambda$ is determined by the electro-neutrality condition and $n_{h}(z)=b\exp\left(-u_{0}(\vecrho,z)\right)$ is the counterion concentration for the equivalent homogeneous surface-charge density, $\langle\sigma(\vecrho)\rangle_{\vecrho}=\sigma_{0}$. For a continuous dielectric ($\eps=\eps'$) across the $z=0$ boundary, the concentration is given by $n_{h}(z)=\left(\sigma_{0}/e\lgc\right)\exp(-z/\lgc)$. As is evident from Eq.~(\ref{eq10}), the effect of heterogeneity is encapsulated in the heterogeneous Boltzmann factor $\exp\left(-u_{1}(\vecrho,z)\right)$ and in the prefactor $\Lambda$.

We consider a simple example of a single-mode charge modulation in the $x$ direction, $\sigma_{1}/\sigma_{0}=\Delta\cos(kx)$, with $|\Delta|<1$, and a continuous dielectric  ( $\eps=\eps'$) across the $z=0$ boundary. The contribution to the interaction potential, $u_{1}$, is then given by $u_{1}=\left(\Delta/k\lgc\right)\cos(kx)\exp(-kz)$. By averaging over $x$, we obatin a zeroth-order modified Bessel function of the first kind,
%%%%%%%%%%%eq%%%%%%%%%%
\be
\label{eq1c}
\langle n(\vecrho,z)\rangle_{\vecrho}=\Lambda n_{h}(z)I_{0}\left(\frac{\Delta\e^{-kz}}{k\lgc}\right).
\ee
%%%%%%%%%%%%%%%%%%%%%%%%%
As $\langle n(\vecrho,z)\rangle_{\vecrho}$ decays to zero  faster than $n_{h}(z)$ in the $z$-direction, more counterions accumulate near the surface, as compared with the homogeneous case ($\sigma=\sigma_{0}$). This result is similar to the counterion-only result within PB theory, and it holds also when additional modulation modes are taken into account~\cite{Jho06}.

Within the SC regime, counterion localization at the surface proximity has been demonstrated in several other classes of surface-charge heterogeneity. Namely, the effect was studied for surfaces with random and disordered surface charge~\cite{Naji14} as well as discrete surface charges~\cite{Moreira02,Pezeshkian12}. A further discussion of such setups lies beyond the scope of this work.

In this section we reviewed the ionic concentrations induced by inhomogeneous surface-charge densities in three different electrostatic models. Far from the surface, the main contribution stems from the lower $k$-modes of the surface-charge modulation. Moreover, surface-charge heterogeneity leads to an enhanced localization of counterions at the surface proximity. The effect originates from the nonlinear coupling of different $k$-modes in the full nonlinear PB theory, and is not captured within the linear DH theory.

 Within the SC regime, appropriate for $\Xi\gg1$,  the localization of counterions at the surface is related to the properties of the exponential function;  due to surface-charge modulation, there are regions across the surface to which the counterions are less attracted and those to which they are more attracted. The counterion contact density,  given by a Boltzmann factor of a single counterion, decreases in those former regions and increases in the latter.
Because the exponential function is convex, the excess contribution of those latter regions tips the scales, resulting in an overall increase in the area-averaged contact density.

%%%%%%%%%%%%%%%%%%%%%%%%%%%%%%%%%%%%%%%%%%%%%%%%%%%%%%%%%%%%%%
\section{Interaction between heterogeneously charged surfaces}
\label{sec5}
%%%%%%%%%%%%%%%%%%%%%%%%%%%%%%%%%%%%%%%%%%%%%%%%%%%%%%%%%%%%%%

The effects of surface-charge heterogeneity on the interaction between a pair of charged surfaces in solution have been investigated in several theoretical works~\cite{Richmond74,Milkavcic94,Rudi06,DanPRE,Holt97,Stankovitch99,Velegol01,RamPRE,Ghosal17,LevinMC}. The general approach is to find the electrostatic potential and relate it to the osmotic pressure between the surfaces, $\Pi$. The pressure $\Pi=p_{\rm{in}}-p_{\rm{out}}$ is the difference between the inner pressure and the outer one that is exerted by the bulk of the solution.

\begin{figure}[ht]
\centering
\includegraphics[width=0.85\columnwidth]{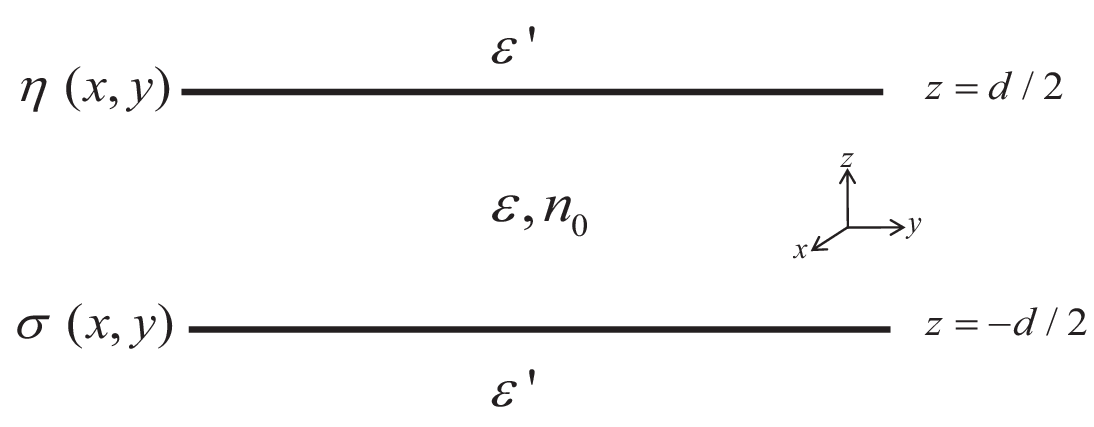}
\caption{Schematic drawing of two aligned planar surfaces separated by a distance $d$ along the $z$-axis. The charge distribution on the top surface is $\eta(x,y)$ and on the bottom surface is $\sigma(x,y)$. The dielectric
constant and salt concentration in the inner region are $\eps$ and $n_{0}$,
respectively. In the outer region, the dielectric constant is $\eps'$ and there are no ions. }
\label{fig6}
\end{figure}

In order to study the osmotic pressure between heterogeneously charged surfaces, consider the following electrostatic setup, as is depicted in Fig.~\ref{fig6}: two parallel and heterogeneously charged planar surfaces are separated along the $z$-axis by a distance $d$. We denote the bottom surface-charge density as $\sigma\,(x,y)$ and the top one as $\eta\,(x,y)$. The surfaces are of area $A$ and separate an inner ionic solution of dielectric constant $\eps$ and bulk ionic concentration $n_{0}$, from an outer medium of dielectric constant $\eps'$, as in Section~\ref{DH_concentrations}.
%%%%%%%%%%%%%%%%%%%%%%%%%%%%%%%%%%%%%%%%%%%%%%%%%%%%%%%%%%%

%%%%%%%%%%%%%%%%%%%%%%%%%%%%%%%%%%%%%%%%%%%%%%%%%%%%%%%%%%%

Once the electrostatic potential in the inner region between the surfaces is calculated, the osmotic pressure can be derived from the Maxwell stress tensor. Including contributions from the \ra{$z{-}z$ component of the} Maxwell stress tensor and the ideal gas pressure of mobile ions yields
%%%eq%%%
\be
\label{eq11}
\Pi= \frac{\eps}{8\pi}\left(E_{x}^{2}+E_{y}^{2}-E_{z}^{2}\right)+\kbt\left(\Delta n_{+}+\Delta n_{-}\right),
\ee
%%%%%%%%%%%%%%%%%%%%%%%%%%%%%%%%%%%%%%%%%%%%%%%%%%%%%%%%%%%%5
where $\Delta n_{\pm}=n_{\pm}(\vecr)-n_{0}$ is the difference in the ionic concentration of the positive/negative ions from that in the bulk, and $\mathbf{E}=\left(E_{x},E_{y},E_{z}\right)$ is the electric field. For homogeneously charged surfaces, the electrostatic potential varies only along the $z$-axis, leading to an attractive term $\sim -E_{z}^{2}$. In the heterogeneous case, on the other hand, the variation of the potential in the $(x,y)$ plane leads to the repulsive terms $\sim \left(E_{x}^2+E_{y}^2\right)$.
As the osmotic pressure in thermodynamic equilibrium is constant throughout the system, it is possible to evaluate Eq.~(\ref{eq11}) at arbitrary positions, such as at the midplane, $z=0$.

The osmotic pressure can also be derived from the free energy via the thermodynamic identity $p_{\rm{in}}=-\left(\partial F/\partial d\right)/A$. This identity yields the inner pressure and the outer one is obtained by taking the limit $d\to\infty$, leading to
%%%eq%%%%
\be
\label{eq12}
\Pi=-\frac{1}{A}\left(\frac{\partial F}{\partial d}-\lim_{d\to\infty}\frac{\partial F}{\partial d}\right).
\ee

Many of the important features of the osmotic pressure are captured within the DH framework, for which the electrostatic potential solves the following linear differential equation:
%%%%%%%%%%eq%%%%%%%
\be
 \label{eq11a}
 \nabla^{2}\psi=\begin{cases}
\kd^{2}\psi & -d/2\le z\le d/2\\
0 & |z|>d/2.
\end{cases}
 \ee
%%%%%%%%%%%%%%%%

The boundary conditions for the potential are defined by the surface-charge densities according to
  %%%%%%%%%%%%%%%%%%%%%
  \begin{align}
  \label{eq13a}
  \left.\eps\dz\psi\right|_{-d/2^{+}}-\left.\eps'\dz\psi\right|_{-d/2^{-}}&=-4\pi\sigma\,(x,y),\nonumber\\ \nonumber\\ \left.\eps\dz\psi\right|_{d/2^{-}}-\left.\eps'\dz\psi\right|_{d/2^{+}}&=4\pi\eta\,(x,y).
  \end{align}
  %%%%%%%%%%%%%%%%%%%

 The DH equation can be readily solved and the osmotic pressure can then be derived. The DH osmotic pressure is most conveniently expressed via the Fourier transform in the in-plane $(x,y)$ coordinates, as was introduced in Section~\ref{DH_concentrations}. One finds that~\cite{Richmond74,Milkavcic94,Rudi06,DanPRE}

%%%%%%%%%%eq%%%%%%%%%%%%%%%%%%
\begin{align}
\label{eq14}
\Pi&=\frac{1}{A}\int\frac{\D^{2}k}{\pi}\frac{\Delta_{k}\e^{-2qd}\left(\sigma_{\veck}\sigma_{-\veck}+\eta_{\veck}\eta_{-\veck}\right)}{\left(1-\Delta_{k}^{2}\e^{-2qd}\right)^{2}}\Gamma_{k}\nonumber\\ &+\frac{1}{A}\int\frac{\D^{2}k}{2\pi}\frac{\left(\e^{-qd}+\Delta_{k}^{2}\e^{-3qd}\right)\left(\sigma_{\veck}\eta_{-\veck}+\eta_{\veck}\sigma_{-\veck}\right)}{\left(1-\Delta_{k}^{2}\e^{-2qd}\right)^{2}}\Gamma_{k}, \end{align}
%%%%%%%%%%%%%%%%%%%%%%%%%%%%%%
where $\Gamma_{k}=q(1+\Delta_{k})/\left(q\eps+k\eps'\right)$, $k=|\veck|$, and $q^{2}=k^{2}+\kd^{2}$. In Eq.~(\ref{eq14}) a new $k$-dependent quantity has been introduced: $\Delta_{k}=\left(q\eps-k\eps'\right)/\left(q\eps+k\eps'\right)$. This parameter corresponds to image interactions induced by the discontinuity of the dielectric constant and ionic concentration at $z=\pm d/2$. Given a large dielectric mismatch, $\eps'\ll\eps$, it is customary to approximate $\Delta_{k}=1$ and $\Gamma_{k}=2/\eps$. Similar calculations were performed in spherical geometry for charged colloidal particles (see, for example, Refs.~\cite{Holt97,Stankovitch99,Velegol01} ).

Note the difference between the two terms of Eq.~(\ref{eq14}). The first term accounts for the self-energies of the two bounding surfaces in presence of the discontinuity of the dielectric constant and the ionic concentration. Therefore, this term has the same sign as $\Delta_{k}$, regardless of the surface-charge densities. In particular, it is repulsive for $\eps>\eps'$. The second term accounts for the interaction between the two bounding surfaces and its sign depends on the surface-charge densities.

Below, we distinguish between three possible electrostatic setups. For {\it overall charged} surfaces, the effect of heterogeneity is smeared out at large inter-surface separations, and the leading contribution stems from the net charge. However, in the case of {\it overall neutral} surfaces, this net contribution vanishes and the leading contribution depends on the inter-surface charge correlations and the largest patch size. Finally, when the surface-charge densities are {\it random and quenched}, an average over different surface-charge density configurations must be performed.  Then, in the absence of correlations, an important role is attributed to the asymmetry between repulsion and attraction in PB theory. This scenario is highly relevant to the experimentally observed attraction between overall neutral patchy surfaces, described in Section~\ref{patchy_forces}.

\subsection{Overall charged surfaces}
\label{net_charge}
%%add the homogeneous results
 In the case of overall charged surfaces, the effect of any inhomogeneities is smeared out at large separations and the interaction depends solely on the net charge. For example, from the DH result of Eq.~(\ref{eq14}), the pressure satisfies $\Pi\sim\sigma_{0}\eta_{0}\exp(-\kd d)$ for $\kd d \gg 1$. This asymptotic behavior holds also beyond the DH framework~\cite{Lukatsky02b,Milkavcic95} and is the main justification in describing surfaces as homogeneously charged.

 It is possible, however, to discuss corrections to this homogeneous result. In the linear DH framework, the correction is a simple additive term, because each surface-charge mode induces an independent osmotic pressure term. The question remains: what is the effect of heterogeneity in the presence of a net surface charge, beyond DH?

 To answer this question, one can, in general, expand the potential according to the partially linearized framework, as is described in Section~\ref{WC_concentration}. In the case of added salt, this framework was employed by Miklavcic~\cite{Milkavcic95} to study how the contribution of a homogeneous surface-charge density affects the contribution of $k>0$ modulations. By solving the leading correction, $\phi_{1}$, and varying the net charge ($k=0$ mode), it was found that the $k>0$ modulations have a smaller contribution to the osmotic pressure for surfaces with a net charge, as compared to overall neutral ones. Note the difference from the DH result in which all terms are independent and add up by the principle of superposition. This finding implies that the influence of surface-charge inhomogeneity is restricted to surfaces that have a zero net charge or close to it.

Alternatively, by varying the $k>0$ modes for a given net charge, it is possible to examine how surface-charge modulations modify the osmotic pressure, as compared to the homogeneous case. Lukatsky et al.~\cite{Lukatsky02b} addressed this problem in the counterions-only case. They considered periodic surface-charge densities with a relative inter-surface displacement $\vecrho'$, {\it i.e.}, $\eta(\vecrho)=\sigma(\vecrho+\vecrho')$ \cite{Lukatsky02b}, at small and large separations for in-phase and out-of-phase configurations ($\sigma=\eta$ and $\sigma=-\eta$, respectively).

 The effect of surface-charge modulation on the osmotic pressure is two-fold. This can be understood from Eq.~(\ref{eq11}), by examining both the Maxwell stress tensor and counterion concentration at the midplane. For all separations and displacements, the $x$ and $y$ components of the electric field increase the osmotic pressure, while the $z$ component reduces it. Due to symmetry, the reduction is largest in the out-of-phase configuration ($\sigma=-\eta$), and vanishes in the in-phase one ($\sigma=\eta$). The midplane counterion concentration, on the other hand, is always reduced. An explanation can be found in Section~\ref{WC_concentration}; the counterion concentration is increased at the surfaces, $z=\pm d/2$, and reduces at the midplane due to conservation of counterions. Similar arguments can be used also for the SC result~\cite{Jho06}.

 At large separations, these contributions lead to a reduction in the osmotic pressure, independent of the inter-surface displacement, scaling as $d^{-3}$, reminiscent of Casimir-type forces. At small separations, the osmotic pressure is always reduced in the out-of-phase configuration, while in the in-phase one, it is reduced for large modulation wavelengths but increases for small ones. Note that in both limits, the pressure is reduced for large modulation wavelengths.

\subsection{Correlations between overall neutral surfaces}
\label{correlations}
%write the two important features: modulation wave vectors and correlations
In the absence of a net charge, other properties of the surface-charge densities, $\sigma$ and $\eta$, come into play. At large inter-surface separations, as mentioned above, the leading contribution to the osmotic pressure stems from the minimal mode $k_{{\rm min}}$, corresponding to the largest patch size. If both $\sigma$ and $\eta$ contain this mode, the inter-surface term $\sim\exp\left(-\qmin d\right)$ is the dominant one. Otherwise, the dominant contribution comes from the self-energy term $\sim\exp\left(-2\qmin d\right)$ that decays twice as fast.

The previous paragraph demonstrates the importance of another feature of the surface charge densities, {\it i.e.}, the surface-charge correlations. The significance of correlations is evident in the DH result of Eq.~(\ref{eq14}), written in terms of these products: $\sigma_{\veck}\sigma_{-\veck}$, $\eta_{\veck}\eta_{-\veck}$, $\sigma_{\veck}\eta_{-\veck}$, and $\eta_{\veck}\sigma_{-\veck}$. These are the Fourier transforms of the two-point auto-correlation and inter-correlation functions, respectively, where the two-point correlation function of $\sigma$ and $\eta$ is defined as
%%%%%%%%%%%%%%%%eq%%%%%%%%%%%5
\be
\label{eq15a}
G_{\sigma,\,\eta}(\vecrho)=\int\D^{2}\vecrho'\,\sigma(\vecrho')\eta(\vecrho'+\vecrho).
\ee
%%%%%%%%%%%%%%%%%%%%%%%%%%%%%%%%%%%%%

\begin{figure} [ht]
\centering
\includegraphics[width=0.7\columnwidth]{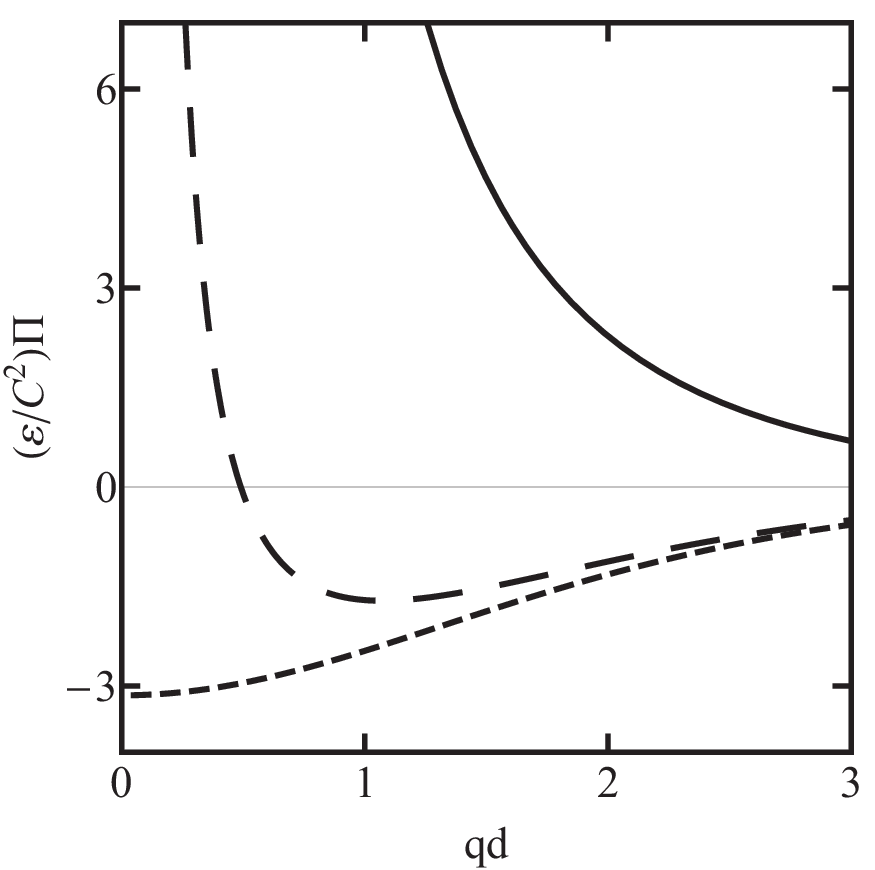}
\caption{Dimensionless pressure profiles, $\left(\eps/C^{2}\right)\Pi$, for a single mode surface-charge modulation with $C_{\sigma}=C_{\eta}=C$. Profiles are plotted for three phases: $\delta=0$ (solid), $\delta=0.85\pi$ (long-dashed) and $\delta=\pi$ (short-dashed). As the phase increases, the inter-surface term becomes attractive until the entire pressure profile becomes negative.}
\label{fig7}
\end{figure}

We further discuss the role of correlations by considering only a single mode of surface-charge modulation in the $x$-direction, $\sigma(\vecrho)=C_{\sigma}\cos(kx)$ and $\eta(\vecrho)=C_{\eta}\cos(kx+\delta)$, for a large dielectric mismatch ($\eps\gg\eps'$). The relative phase, $0\le\delta<\pi$, determines the inter-surface correlation, and Eq.~(\ref{eq14}) reduces to~\cite{DanPRE}
%%%%%%%%%%%%%%%%%%%%%%%%%%%%%%%%%%%%
\be
\label{eq15}
\Pi=\frac{\pi}{\eps}\frac{C_{\sigma}^{2}+C_{\eta}^2+2C_{\sigma}C_{\eta}\cosh qd\cos\delta}{\sinh^{2}qd}.
\ee
%%%%%%%%%%%%%%%%%%%%%%%%%%%%%%%%%%
For phases in the range $\delta<\pi/2$, the correlation term adds to the self-energy repulsion. In the range $\pi/2<\delta<\pi$, on the other hand,
the correlation term is negative. It competes with the self-energy repulsion and becomes dominant at large separations. In this case, the osmotic pressure turns over from repulsive to attractive with increasing separations, and the crossover separation, $d^{\ast}$, is given by
%%%%%%%%eq%%%%%%%%%%
\be
\label{eq15b}
qd^{\ast}=\cosh^{-1}\frac{C_{\sigma}^{2}+C_{\eta}^{2}}{2C_{\sigma}C_{\eta}|\cos\delta|}.
\ee
%%%%%%%%%%%%%%%%%%%%%%%%
 Pressure profiles for different values of the phase $\delta$ are illustrated in Fig.~\ref{fig7}.

%%%%%fig%%%%%%%%%

In the special case of $\delta=\pi/2$, the correlation term, $\sim\cos\delta$, in Eq.~(\ref{eq15}) vanishes. At large separations, the osmotic pressure decays as $\sim\exp\left(-2qd\right)$, which is a general feature of the DH result between surfaces with no correlation. As the pressure decays twice as fast than it does with correlations, the corrections to the DH calculation become important in this case even for small potentials \cite{RamPRE}. This fact is demonstrated in Section~\ref{pathcy_theory}, where the pressure is calculated between surfaces with random patch arrangements and inter-surface correlations that vanish on average.

\subsection{Overall neutral and randomly charged surfaces}
\label{pathcy_theory}
In many experimental setups, the heterogeneous surface-charge densities cannot be determined exactly. For example, two surfaces that are charged using the same experimental procedure, may exhibit some unknown relative displacement, $\eta(\vecrho)=\sigma(\vecrho+\vecrho')$. Furthermore, the charging process itself may be partially random in nature. This is the case with the preparation process of patchy surfaces, as is described in Section~\ref{preparation}, where patches of positive surfactant bilayers form on negative mica surfaces in an uncontrolled arrangement. For such random surface-charge densities,  the nature of the interaction depends greatly on whether the system is {\it annealed} or {\it quenched}. In the former, the random surface-charge density is in thermodynamic equilibrium, while in the latter case it is ``frozen" in time.

Correlations are omnipresent in annealed systems, where the surface charges are mobile and can lock into thermodynamically favorable configurations. In the case of overall neutral systems, for example, surface charges rearrange themselves such that the two surfaces are oppositely charged ($\eta=-\sigma$), resulting in a lower free energy. The surface-charge density  ($\sigma$ or $\eta$) can then be determined by the interplay between electrostatics and other short-range interactions~\cite{Brewster08,Jho11}, as is described in Section~\ref{sec3}. An interesting role was suggested for salt in this situation; for higher salt concentrations, the suppressed electrostatics allow line tension to induce larger surface-patches, possibly resulting in a stronger electrostatic attraction~\cite{Brewster08}. MC simulations, however, support the traditional role of salt~\cite{Jho11}. Although the patch size increases for higher ionic concentrations, the screened electrostatics results in an overall weaker attraction.

In the second scenario, random and quenched systems exhibit no inter-surface correlations on average. The correlation term of the DH osmotic pressure $\sim\sigma_{\veck}\eta_{\veck}$  vanishes, yielding an overall repulsion. Furthermore, for ``molecular-size patches", Podgornik and Naji~\cite{Rudi06} have shown that DH theory predicts a repulsive interaction even when incorporating fluctuation effects beyond MF in a loop expansion of the free energy. In light of this result, the long-range attraction between quenched patchy surfaces, as measured by Silbert et al.~\cite{SilbertPRL}, is surprising and must originate from another source.

\begin{figure}[ht]
\centering
\begin{subfigure}[b]{1\columnwidth}
\centering
\includegraphics[width=0.6\textwidth]{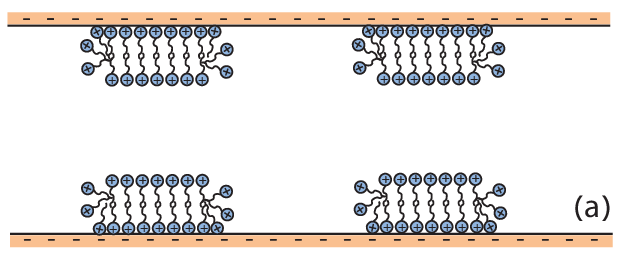}
%\caption{}
\end{subfigure}
\begin{subfigure}[b]{1\columnwidth}
\centering
\includegraphics[width=0.6\textwidth]{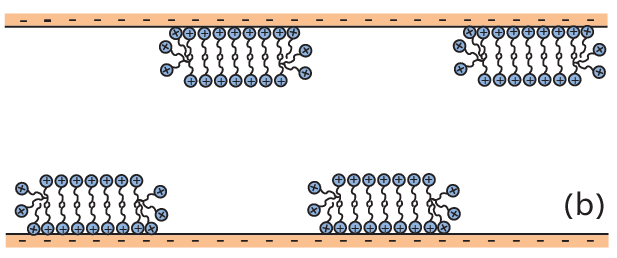}
%\caption{}
\end{subfigure}
\caption{(Color online) Two patchy surfaces in (a) fully  commensurate and (b) fully incommensurate configurations. In Ref.~\cite{SilbertPRL}, the random patch arrangement is considered as an idealized combination of these two configurations. The osmotic pressure in both configurations is approximated as the pressure between two homogeneously charged surfaces, either equally or oppositely charged, respectively. The total osmotic pressure is the average between the two.}
\label{fig8}
\end{figure}

One possible explanation lies the nonlinear corrections appearing in the full non-linear PB theory for finite size patches. Together with the experimental evidence of attraction, Silbert et al.~\cite{SilbertPRL} have also suggested a possible theoretical explanation for attraction using a simple averaging argument. The osmotic pressure between randomly arranged patches is approximated by an average over two situations of interaction between two infinite and homogeneously charged surfaces. In the first, the surfaces are equally charged, while in the second, the surfaces are oppositely charged. Thus, each charged patch, considered to be very large, faces with equal probability either an equally charged patch or an oppositely charged one, as is illustrated in Fig.~\ref{fig8}.  Their numerical calculation of $\Pi=\left(\Pi_{+/+}+\Pi_{+/-}\right)/2$, where $\Pi_{+/+}$ is the osmotic pressure between equally charged surfaces and $\Pi_{+/-}$ between oppositely charged ones, shows that the attraction in the latter case ($\Pi_{+/-}$ ) exceeds the repulsion in the former one ($\Pi_{+/+}$), yielding an overall attraction within the nonlinear PB theory.

%%%%%%%%%%%%%%%%%%%%%%patchy configurations%%%%%%%%%%%%%%%%%%%%%%%%%%%%%%%
%%%%%%%%%%%%%%%%%%%%%%%%%%%%%%%%%%%%%%%%%%%%%%%%%%%%%%%%%%%%%

The difference between repulsion and attraction can be captured analytically by examining the full non-linear PB equation [Eq.~(\ref{eq5a})] for homogeneously charged surfaces. The first to investigate the osmotic pressure between asymmetrically charged surfaces were Parsegian and Gingell~\cite{Parsegian72}, who addressed the scenario of added electrolyte and obtained the criteria for repulsion and attraction within the DH framework. Such criteria were later extended for the counterions-only case~\cite{Lau99} and for the general nonlinear PB framework~\cite{Dan07}. In particular, two studies were dedicated to the interaction between oppositely charged surfaces \cite{MeierKoll04,SamEPL}, demonstrating the significance of counterion release. It was shown that for large surface-charge densities and small salt concentrations, counterions between oppositely charged surfaces are released into the bulk due to entropic gain, enhancing the electrostatic attraction. Explicitly, under these conditions and at separations $\lgc\ll d\ll\ld$, the attraction between oppositely charged surfaces scales as $\Pi_{+/-}\sim-d^{-2}\ln^{2}\left(d/8\ld\right)$~\cite{SamEPL}, as opposed to the repulsion between equally charged surfaces that scales as $\Pi_{+/+}\sim d^{-2}$. Therefore, the ratio between the two satisfies
%%%%%%%%%%eq%%%%%%%%%
\begin{equation}
\label{eq16}
\left|\frac{\Pi_{+/-}}{\Pi_{+/+}}\right|\sim\ln^{2}\frac{d}{8\ld}.
\end{equation}
%%%%%%%%%%%%%%%%%%%%

The forces between patchy surfaces were recently tested in MC simulations~\cite{LevinMC} for strongly charged patches and weak ionic strength. In the simulations, surfaces of area $\sim1,500\,{\rm nm}^{2}$ were divided into two or four patches, charged alternatively with positive or negative charges. Similarly to the averaging framework described above, the net interaction between the surfaces in the simulation was evaluated as the average over the different configurations of two or four patches. An attraction was found in both cases and was stronger for the larger patches~\cite{LevinMC}. Moreover, the attraction was weaker than the one predicted by an average over two homogeneous systems as is described above.

For an arbitrary patch size and weak electrostatic interaction, an analytic expression can be derived for the osmotic pressure between patchy surfaces~\cite{RamPRE}, as is reviewed next. Consider the electrostatic setup of Section~\ref{sec5}. For simplicity, assume that the electric field is confined within the inner medium
 ($n'=\eps'=0$) and that the surface-charge densities have a form $\sigma=\sigma_{k}\cos(kx)$ and $\eta=\sigma_{k}\cos(kx+\delta)$, corresponding to patchy stripes of width, $w=\pi/k$, common to both surfaces. The relative phase, $\delta$, is arbitrary and depends on the specific experimental setup. Assuming that the surfaces are sufficiently large, every possible value of $\delta$ should be manifested. Therefore, the calculation of physical quantities requires an average over the phase, $\delta$. Given that the surfaces were prepared separately, $\delta$ is likely to be distributed uniformly in the range $0\le\delta\le\pi$.

In this electrostatic setup, within the DH framework, the screening length is given by $q^{-1}=1/\sqrt{k^{2}+\kd^{2}}$ as is explained in Section~\ref{DH_concentrations}. The dimensionless factor $\left(q\lgc\right)^{-1}$ can be used to characterize the strength of the electrostatic interaction, and it diminishes for a combination of small surface charge and screening-length.  For $q\lgc>1$, it is possible to expand the dimensionless electrostatic potential, $\phi$, in powers of the small parameter $\left(q\lgc\right)^{-1}$, according to $\phi=\left(q\lgc\right)^{-1}\phi_{1}+\left(q\lgc\right)^{-3}\phi_{3}+\mathcal{O}\left[(q\lgc)^{-5}\right]$. The expansion contains only odd powers as the potential is odd in the surface-charge density, $\sigma\sim\lgc^{-1}$. In particular, the DH potential is obtained as the first-order term in the expansion. In Ref.~\cite{RamPRE}, for $q\lgc>1$, the electrostatic potential was approximated by the first two terms, by applying a variational principle to the PB free energy. The resulting osmotic pressure was averaged over a uniform distribution of the phase, $0\le\delta\le\pi$.
%%%%%%%%%% figure 8: pressure profiles from our PRE %%%%%%%%%%%
\begin{figure*}[ht]
\centering
\begin{subfigure}[b]{0.45\textwidth}
\includegraphics[width=0.85\textwidth]{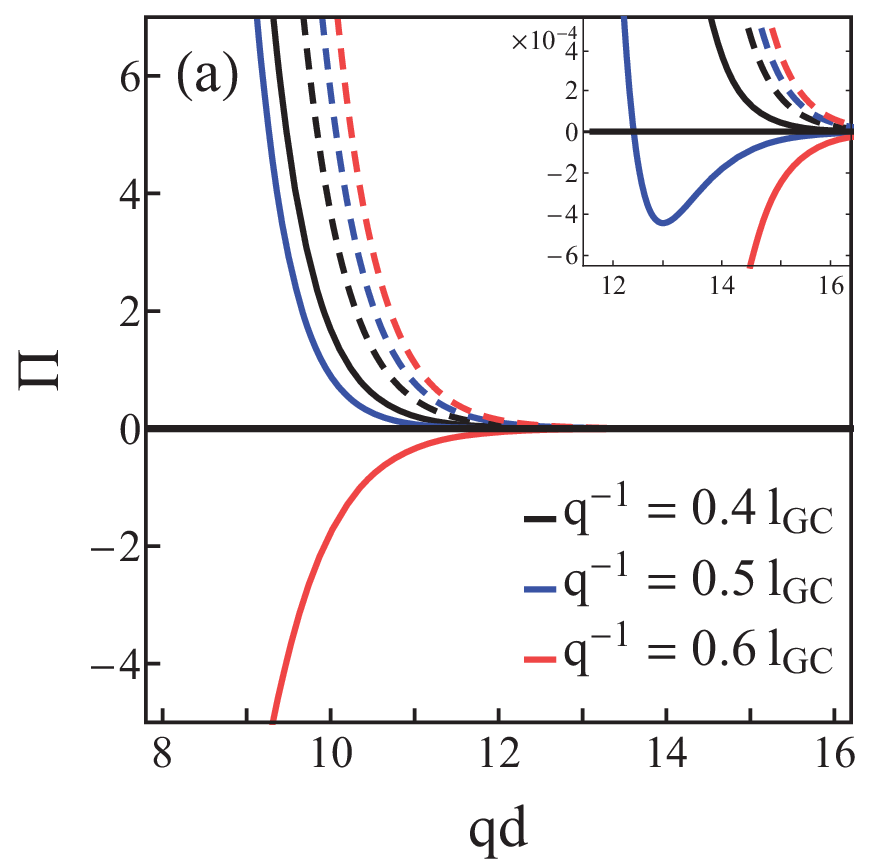}
\end{subfigure}
\begin{subfigure}[b]{0.45\textwidth}
\includegraphics[width=0.85\textwidth]{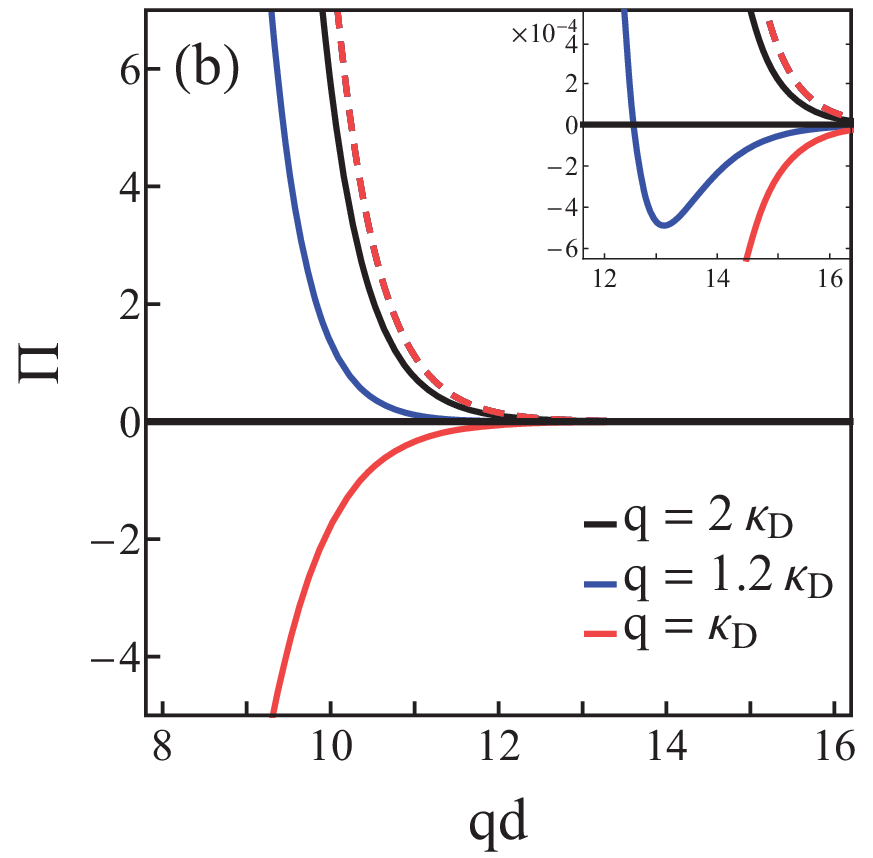}
\end{subfigure}
\caption{(Color online) Osmotic pressure profiles in units of $10^{-9}\,\kbt q^{2}/\left(2\pi\lb\right)$ as function of $qd$,  between patchy surfaces according to the expansion in powers of $\left(q\lgc\right)^{-1}$. For $T=298\,{\rm K}$, $\eps=80$, and $q^{-1}=1\,{\rm nm}$, the osmotic pressure is in units of mPa. The DH result ($1^{\rm{st}}$-order term) is plotted as dashed curves and the next order (sum of $1^{\rm{st}}$-order and $3^{\rm{rd}}$-order terms) is plotted as solid ones, demonstrating the limitation of the DH approximation for  uncorrelated surfaces. (a) Pressure profiles between surfaces with infinitely large charged patches for three patch-charge densities. (b) Pressure profiles between surfaces with a fixed patch charge-density for three patch width values. In both (a) and (b), the intermediate profile (blue) crosses over from repulsion to attraction at smaller pressure values, as is shown in the corresponding insets. Adapted from Ref.~\cite{RamPRE}.}
\label{fig9}
\end{figure*}
%%%%%%%%%%%%%%%%%%%%%%%%%%%%
At large inter-surface separations ($qd\gg 1$), the osmotic pressure is given by:
%%%%%%%%%%%%%%%%eq%%%%%%%%%%%%%%%%%%%%%%%
\begin{align}
\label{eq17}
\Pi & \approx\frac{2\kbt}{\pi\lb\lgc^2}\left(1-\frac{d}{d^{\ast}}\right)\e^{-2qd},
\end{align}
%%%%%%%%%%%%%%%%%%%%%%%%%%%%%%%%%%%%%%%%%%%%%
where $d^{\ast}=4q^{3}\left(\ld\lgc\right)^{2}$ is the crossover separation, at which the pressure crosses over from repulsive to attractive. Results for the osmotic pressure are depicted in Fig.~\ref{fig9}.

 As is evident from the expression for $d^{\ast}$, a crossover occurs for all possible values of $q$, $\ld$ and $\lgc$, {\it i.e.}, attraction always prevails at large separations. The first correction to the DH potential, as small as $\left(q\lgc\right)^{-3}$, suffices to yield an overall attraction. An explanation is found in Section~\ref{correlations}. For a given phase, $\delta$, the correction to DH at large separations is negligible as compared to the leading correlation term $\sim\exp\left(-qd\right)$. However, as the average over the phase $\delta$ cancels out these terms, only the self-surface terms $\sim\exp\left(-2qd\right)$ are left and the correction becomes important. This is an example of the limitation of the DH theory to describe the forces between overall neutral surfaces with no inter-surface correlations, even for large inter-surface separations.

 We note that the three above frameworks provide insight into the attraction between two overall neutral patchy surfaces, but a limited one. The averaging scheme of Ref.~\cite{SilbertPRL} conveys the asymmetry between repulsion and attraction within PB theory, but it is valid only for $\lgc\ll d\ll \ld\ll w$, and yields an osmotic pressure that is insensitive to $\lgc$ and $w$. The MC simulations of Ref.~\cite{LevinMC} enable the study of a wide range of physical conditions, but only a few of them have been tested. In particular, it would be interesting to perform such simulations for more random surface-charge densities. Finally, the framework of Ref.~\cite{RamPRE} describes the dependence of the osmotic pressure on all the system lengthscales, but it does so only for weak interactions.

\section{Comparison to van der Waals attraction}
\label{sec6}
% read and address the papaers by Naji on fluctuation induced interactions.
It is instructive to compare the electrostatic interaction between overall neutral \ra{quenched} patchy surfaces with the ever-present van der Waals (vdW) attraction between uncharged surfaces. While the first originates from the averaged electrostatics between surface-charge patches and depends on the specific form of the surface-charge density, the latter stems from correlated dipole fluctuations, existing between any two surfaces. Despite their different origins, the two interactions have comparable magnitudes at large inter-surface separations.

In the presence of salt, the zero-frequency vdW attraction decays exponentially rather than algebraically~\cite{ParsegianVDW,NinhamVDW,NetzVDW}. In the limit of large separations ($\kd d\gg 1$), the vdW force per unit area, $f_{\rm{vdW}}$, is given by \cite{NinhamVDW}
%%%%%%%%%%%%%eq%%%%%%%%%%%%%%%%
\be
\label{eq18}
f_{\rm{vdW}}=-\frac{\kbt\kd^3}{4\pi}\frac{\e^{-2\kd d}}{\kd d},
\ee
%%%%%%%%%%%%%%%%%%%%%%%%%%%%%%%
independent of any surface properties. \ra{We emphasize that $f_{\rm{vdW}}$ accounts for the correlated dipolar fluctuations and not for possible monopolar charge fluctuations that can also be considered beyond MF~\cite{Rudi98,Naji10}.}

 Within the DH framework for molecular-size patches, the repulsive electrostatic interaction is proportional to the above expression and effectively renormalizes the vdW force~\cite{Rudi06}. This result can be obtained by averaging Eq.~(\ref{eq14}) over surface-charge densities with no inter-surface correlations. The surface-charge densities are taken to be randomly distributed according to a Gaussian distribution $\langle\sigma(\vecrho)\sigma(\vecrho')\rangle_{\sigma}=\langle\eta(\vecrho)\eta(\vecrho')\rangle_{\eta}=\gamma^{2}a^{2}\delta\left(\vecrho-\vecrho'\right)$, where $\gamma$ is the root-mean-square surface-charge density, and $a$ is a conveniently defined microscopic length~\cite{Rudi06,DanPRE}. Then, assuming no ions in the outer region, the osmotic pressure at large inter-surface separations is given by
%%%%%%%%%%%%%%%eq%%%%%%%%%%%%%%%%%%%%%
\be
\label{eq19}
\Pi=\frac{4\gamma^{2}a^{2}\kd}{\eps}\frac{\e^{-2\kd d}}{d}.
\ee
%%%%%%%%%%%%%%%%%%%%%%%%%%%%%%%%%%%%%%%

Going beyond DH, as is described in Section~\ref{pathcy_theory}, the vdW and electrostatic interactions can be comparable at large separations only if they have comparable screening lengths, $q^{-1}\approx\ld$, corresponding to very large patches ($w\gg\ld$). Comparing Eqs.~(\ref{eq17}) and (\ref{eq18}) for such large patches, one finds that the electrostatic attraction is dominant for
%%%%%%eq%%%%%%%%%
\be
\label{eq20}
\left(\frac{\pi\lgc}{w}\right)^{4}\,<\,2\frac{\ld}{\lb} u^{2}\e^{-u},
\ee
%%%%%%%%%%%%%%%%%
where $u\equiv d\ld/ w^{2}$ is a dimensionless parameter. The ratio on the left-hand side of Eq.~(\ref{eq20}) is inversely proportional to the total patch charge, while the ratio on the right-hand side depends solely on bulk properties and increases with the salt concentration.  The function $f(u)=u^{2}\e^{-u}$ is bounded from above by about $0.5$, implying that the long-range electrostatics are comparable with vdW only for $\left(\pi\lgc/w\right)^{4}<\ld/\lb$. Under reasonable physical conditions, the electrostatic term is dominant over a wide range of separations. For example, for $T=300\,\rm{K}$, $n_{0}=2\,\rm{mM}$,  $e/\left(|\sigma|w\right)=3\,\rm{nm}$, and $w=100\,\rm{nm}$, the electrostatic term is dominant for separations up to $d=650\,\rm{nm}$~\cite{RamPRE}.

Unlike the zero-frequency vdW attraction,  higher frequency terms are not affected by the presence of salt ions~\cite{NinhamVDW,NetzVDW}. It is possible to incorporate these contributions in the Hamaker constant as in $f_{\rm{vdW}}\sim-\mathcal{H}d^{-3}$, and the Hamaker constant, $\mathcal{H}$, can be calculated using Lifshitz theory~\cite{Parsegian72,Safinyabook}.  The exponentially decaying electrostatic attraction of Eq.~(\ref{eq17}) is thus weaker than the vdW term unless $\kbt\ld/\left(\mathcal{H}\lb\right)\gg1$, corresponding to a combination of small ionic concentrations and small Hamaker constant. Note that the Hamaker constant does not contain the zero-frequency contribution.

%%review dynamic vdW again

%%%%%%%%%%%%%%%%%%%%%%%%%%%%%
\section{Summary and outlook}
\label{outlook}
%%%%%%%%%%%%%%%%%%%%%%%%%%%%%

In this paper, we reviewed experimental and theoretical studies of surface-charge inhomogeneity across aqueous ionic solutions. Motivated by recent SFA experiments on thin mica sheets coated with surfactants,
we focused on modeling planar surfaces with charge modulations over mesoscopic length scales. Note that several works have been devoted  in recent years also to the study of heterogeneously charged cylindrical and spherical colloids, as well as to randomly charged surfaces, manifesting heterogeneity on the microscopic level.%%%%add reff

We have discussed under what conditions charged surface patches of finite size can form spontaneously and remain stable on surfaces immersed in an aqueous solution. The surface-charge heterogeneity then results in a {\it universal increment} in the counterion concentration at the surface proximity, which can consequently lead to a reduction in the osmotic pressure between two surfaces.

We have demonstrated that the interaction between overall charged surfaces is mostly determined by their net charge. For overall neutral surfaces, the interaction strength depends on the minimal wavenumber (largest wavelength) of the surface-charge modulation and on inter-surface correlations. The sign of the interaction (attractive/repulsive) depends mostly on inter-surface correlations and on the dielectric discontinuity at the bounding surfaces.

In the scenario of overall neutral patchy surfaces with random patch arrangement, we have discussed the differences between the attraction of {\it annealed} surfaces and {\it quenched} ones. While annealed surfaces can rearrange their charge distributions and exhibit a clearly understood electrostatic attraction between oppositely charged surfaces, quenched surfaces can attract one another because of nonlinear terms in the PB theory, which favor attraction over repulsion.

The asymmetry between repulsion and attraction is currently attributed to counterion release into the bulk. This mechanism is made possible by the overall neutrality of the surfaces even for homogeneous surfaces. As heterogeneous patchy surfaces are more complex, a more elaborate mechanism may play a leading role. In future works, it will be worthwhile to explore in more detail how the surface heterogeneity propagates into the ionic solution bounded by the surfaces. In other words, one should ask not only {\it how many} counterions there are, but also {\it how} they are distributed in between the surfaces.

%A first step towards a more complete description can be to test the osmotic pressure and 3D ionic distribution between patchy surfaces via Monte Carlo (MC) and molecular dynamics (MD) simulations (see, for example, Ref.~\cite{JoePPM}). To date, such simulations have been used between planar patchy surfaces only to investigate the osmotic pressure only for relatively strong interactions. More thorough simulations are expected to be time consuming due to the large number of ions in the solution, which should be treated explicitly in order to obtain 3D distributions. Eventually, it will be of great interest and importance to explain the effect analytically under general physical conditions. As it involves solving non-linear differential equations and averaging the results over a set of random boundary conditions, such a theory will entail a considerable challenge.

The current understanding of long-range attraction between quenched patchy surfaces is rather incomplete. The theoretical frameworks reviewed in this paper account for the phenomenon only in the complementary limits of weak or strong interactions. A thorough description of the system at hand over a complete set of physical parameters, including different possible manifestations of heterogeneity, remains to be established.
%%%%%%%%%%%%%%%%%%%%%%%%%%%%%%%%%%%%%%%%%%%%%%%%%%%%%%%%%%%%%%%%%%%%%%%%%
\section*{Acknowledgments}
%%%%%%%%%%%%%%%%%%%%%%%%%%%%%%%%%%%%%%%%%%%%%%%%%%%%%%%%%%%%%%%%%%%%%%%%%
It is our pleasure and honor to dedicate this article to Dominique Langevin, with whom we had the fortune to collaborate in the past. Dominique greatly contributed to our understanding of fluid interfaces and the dynamics of Langmuir monolayers, and in more general terms, to colloid and interface science.

This work was supported by the Israel Science Foundation (ISF) under Grant No. 438/12, the U.S.- Israel Binational Science Foundation (BSF) under Grant No. 2012/060, and the ISF-NSFC joint research program under Grant No. 885/15. D.A. would like to thank the hospitality of the IPhT at CEA-Saclay and the Free University Berlin (FUB), where this work has been completed. He acknowledges financial support from the French CNRS and the Alexander von Humboldt Foundation through a Humboldt research award.

%\section*{References}

\end{document}